\documentstyle[12pt,epsf,epsfig,wrapfig]{article}
\begin{document}
\begin{center}

\def\bea{\begin{eqnarray}}
\def\eea{\end{eqnarray}}

{\bfseries 
ABOUT THE $Q^2$ DEPENDENCE OF THE 
ASYMMETRY $A_1$ FROM THE SIMILARITY OF THE $g_1$ AND $F_3$
STRUCTURE FUNCTIONS}
\vskip 5mm
A.V. KOTIKOV$^{1 \dag}$ and D.V. PESHEKHONOV$^{{2} \dag}$
\vskip 5mm
{\small
(1) {\it Bogoliubov
 Laboratory of Theoretical Physics, JINR,  Dubna, 141980 Russia
}
\\
(2) {\it
 Laboratory of Particle Physics, JINR, Dubna, 141980 Russia
}
\\
$\dag$ {\it
E-mail: kotikov@thsun1.jinr.ru, peshehon@sunse.jinr.ru
}}
\end{center}
\vskip 5mm

\begin{abstract}
We consider a new approach for taking into account the $Q^2$ dependence of
asymmetry $A_1$.
This approach is based on the similarity of the $Q^2$ behavior 
and the shape of 
the spin-dependent structure function $g_1(x,Q^2)$ and spin averaged 
structure function $F_3(x,Q^2)$. 

\end{abstract}

\vskip 10mm


An experimental study of the nucleon spin structure is realized \cite{AEL}
by measuring of the asymmetry $A_1(x,Q^2) = g_1(x,Q^2) / F_1(x,Q^2)$.
The most known theoretical predictions on spin dependent structure
function (SF) $g_1(x,Q^2)$ of the nucleon were done by Bjorken \cite{Bj} 
and Ellis and Jaffe \cite{EJ} for the so called {\it first moment value}
$\Gamma_1 = \int_0^1 g_1(x) dx$.

Studying the properties of $g_1(x,Q^2)$ and 
the calculation of the $\Gamma_1$ value require the knowledge of SF
function $g_1$ at the same $Q^2$ in the hole $x$ range.
Experimentally asymmetry $A_1$ is measuring at different values of $Q^2$
for different $x$ bins.
An accuracy of the
modern experiments \cite{EG}-\cite{Q2E154}
allows to analyze data in the assumption \cite{EK93}
that asymmetry $A_1(x)$ is $Q^2$
independent (i.e. 
SF $g_1$ and $F_1$ have the same $Q^2$ dependence)
\begin{eqnarray}
\hspace{8cm} A_1(x,Q^2) = A_1(x) \label{a1}
\end{eqnarray}
\begin{wrapfigure}{l}{8cm}
\vskip -1.2cm 
\epsfig{figure=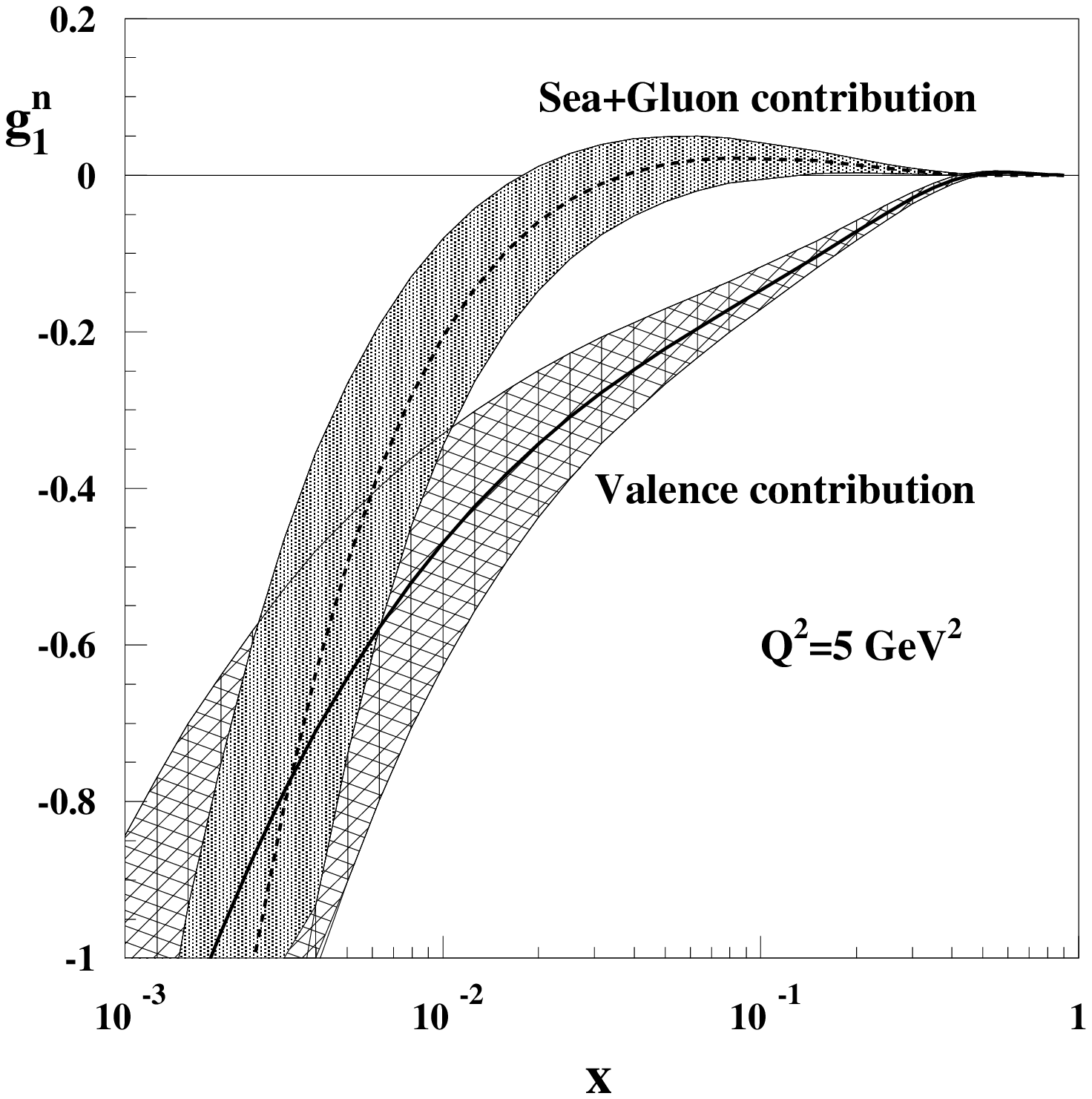,width=7.8cm,height=6cm}
{\small Figure 1: 
The contributions to the structure function $g_1^n$ of the
neutron from the valence quarks
and from the sea quarks and gluons 
The shaded and hatched areas represent the total uncertainties on each
quantity.}
\vskip -0.5cm 
\end{wrapfigure}
But the precise checking of the Bjorken and Ellis - Jaffe sum
rules requires considering the $Q^2$ dependence of $A_1$ or $g_1$.
Moreover, the assumption (1) 
does not  follow from the theory. 
On the contrary, the 
behavior of $F_1$ and $g_1$ as a functions of $Q^2$ is expected to be 
different due to the difference between polarized and unpolarized splitting 
functions.

There are several approaches (see \cite{GRSV}   
and references therein) 
 how to take into account 
the $Q^2$ dependence of $A_1$. They are
based on different approximate solutions of the DGLAP equations.
Some of them have been used already by Spin Muon Collaboration (SMC)
and E154 Collaboration in the last analyses of experimental data
(see \cite{SMC} and \cite{Q2E154}, respectively). \\

\begin{wrapfigure}{l}{8cm}
\vskip 2.2cm 
\hskip 1.0cm
\begin{tabular}{c}
\begin{minipage}[h]{5.cm}
\epsfig{figure=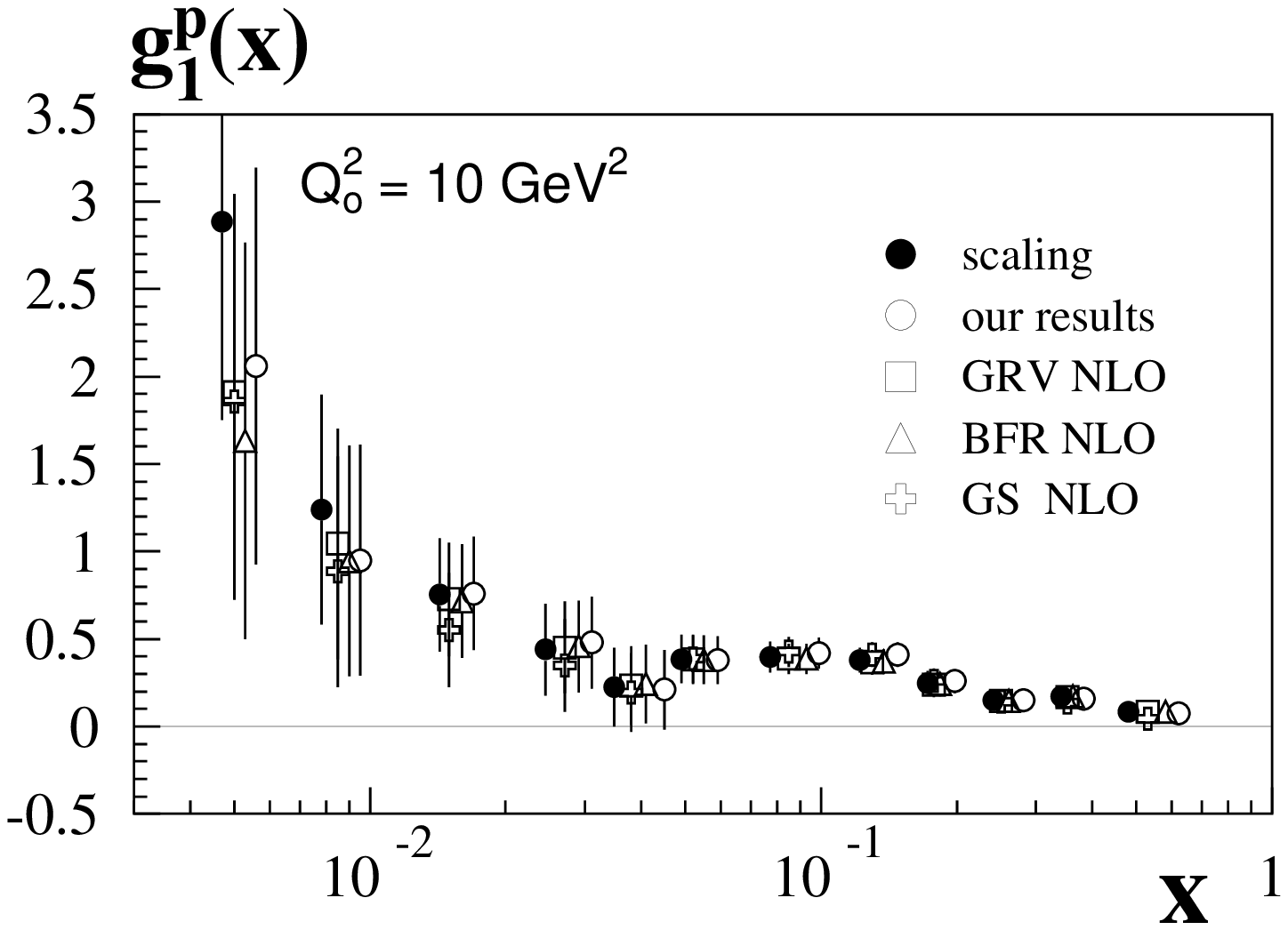,width=6cm}
\end{minipage}
\end{tabular}
\vskip -2.4cm 
{\small Figure 2: The structure function $xg_1^p(x,Q^2)$ evolved to 
$Q^2=10GeV^2$ using our Eq.(2), the assumption that $g_1^p/F_1^p$
is $Q^2$ independent, and exact DGLAP next-to-leading evolution.} 
\vskip -0.3cm 
\end{wrapfigure}

In this paper we review
another idea which is based on the observation in \cite{KP,KNP}
that the splitting functions of  the DGLAP equations for
the $g_1$ and  $F_3$ and  the SF shapes 
are similar in a wide 
$x$ range and, thus, their $Q^2$ 
dependences 
have to be close as a consequence. 
Our approach for $Q^2$-dependence of $A_1$ is very simple 
(see Eq.(\ref{5})) and leads to the results, which are very similar 
to ones based on the  DGLAP evolution.

To demonstrate the validity of the observation, we note that
splitting functions in the r.h.s. of DGLAP equations for
nonsinglet (NS) parts of $g_1$ and $F_3$ are the same
(at least in first two orders of the perturbative QCD)
and differ from $F_1$ one already in the first subleading order.
For the singlet (S) part of $g_1$ and for SF $F_3$ the difference between
splitting functions
is also negligible
(see \cite{KNP}).
This observation allows us to conclude that the ratio of the SF $g_1$ and 
$F_3$ should be essentially
$Q^2$ independent, i.e. 
\begin{eqnarray}
A_1^*(x,Q^2) = {g_1(x,Q^2) \over F_3(x,Q^2)} = A_1^*(x)
\label{a2}
\end{eqnarray}
and
the asymmetry $A_1$ at some  $Q^2$ can be defined than as :
\begin{eqnarray}
A_1(x_i,Q^2) =  {F_3(x_i,Q^2) \over F_3(x_i,Q^2_i)} \cdot
{F_1(x_i,Q^2_i) \over F_1(x_i,Q^2)} \cdot A_1(x_i,Q^2_i),
\label{5}
\end{eqnarray}
where $x_i$ ($Q^2_i$) means an experimentally measured value of $x$ ($Q^2$).

In principle, $Q^2$ independence of the $A_1^*(x)$ ratio can be violated 
at small $x$ values, because the SF $F_3(x)$ has only 
nonsinglet
structure and the SF $g_1(x)$ contains gluon and sea quark contributions.
We 
note, however, that
\begin{itemize}
\item
in polarized case the gluon and sea quark contributions are not so large
even at modern small $x$ values (see Fig. 1 which is taken from 
\cite{Q2E154}): only at $x \leq 10^{-3}$ these contributions start to be 
dominant.
\item
The double-logarithmic 
estimations of small $x$ asymptotics for the SF $g_1$ and  $F_3$  
$g_1^{NS}(x),~F_3(x) \sim x^{-a_{NS}}$ and 
$g_1^{S}(x) \sim x^{-a_{S}}$,
given in 
\cite{Ermolaev},
lead to the results:
$a_{NS} \sim 0.4$ and
$a_{S} \sim 3/2 a_{NS} \sim 0.6$,
that supports a similarity of nonsinglet and singlet polarized components.
\end{itemize}
We 
note that the estimations
in 
\cite{Ermolaev} have been given at small $Q^2$ values.
However, as it was
shown earlier in
\cite{LoYn} and \cite{Ko}, respectively,
the values of $a_{NS}$ and $a_{S}$
should be nearly $Q^2$-independent, 
that is supported also 
by recent fits (see 
\cite{KPS1} and references therein). \\

To apply the proposed approach we use the 
SMC \cite{SMC} 
data (the SLAC \cite{E154n,Q2E154} data have been considered in 
\cite{KNP}).
We parameterize CCFR data on 
$F_2(x,Q^2)$ and $xF_3(x,Q^2)$ 
\cite{CCFRN} and 
 take also the 
SLAC parameterization of 
$R(x,Q^2)$ \cite{SLAC} to obtain structure function $F_1(x,Q^2)$. 
We use in Eq.(\ref{5}) parameterizations of CCFR data \cite{CCFRN} for both 
SF $xF_3(x,Q^2)$ and $F_2(x,Q^2)$ to avoid systematic uncertainties and
nucleon correlation in nuclei.

The SF $g_1(x,Q^2)$ is calculated using the asymmetry $A_1(Q^2)$ as
\begin{eqnarray}
g_1(x,Q^2)= A_1(x,Q^2)\cdot F_1(x,Q^2), 
\label{5.2}
\end{eqnarray}
where the spin average SF $F_1$ has been calculated using NMC 
parameterization of $F_2(x,Q^2)$ \cite{NMC}.
The results are presented in 
Fig. 2.
Our results are in very good
agreement with the calculations
based on direct DGLAP evolution.

\begin{wrapfigure}{l}{8cm}
\vskip -7.5cm 
\hskip -4.2cm
\begin{tabular}{c}
\begin{minipage}[h]{5.cm}
\epsfig{figure=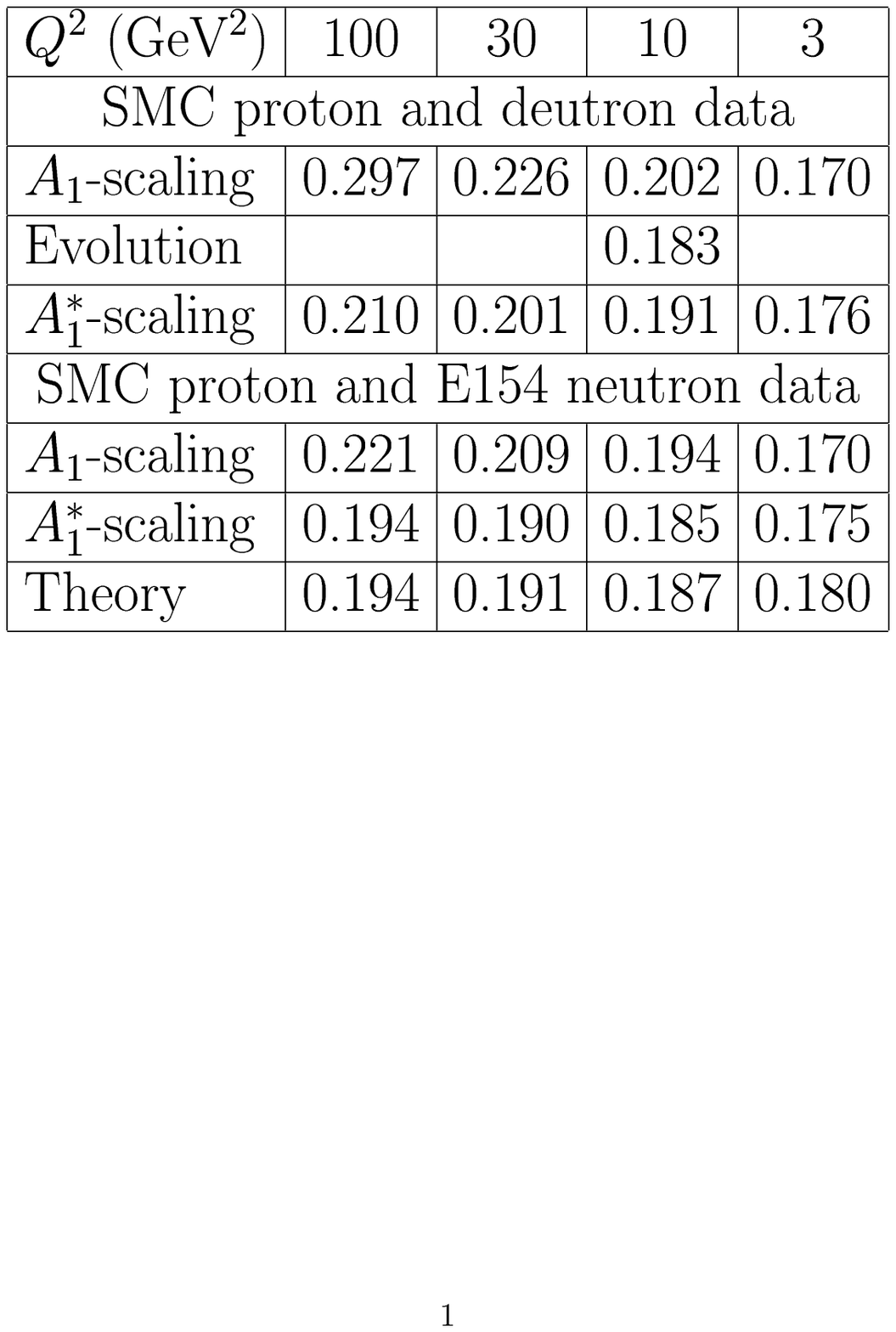,width=14.5cm}
\end{minipage}
\end{tabular}
\vskip -7.2cm 
{\small Table: The mean values of $\Gamma_1^p - \Gamma_1^n $.}
\vskip -0.3cm 
\end{wrapfigure}

To make another comparison with the theory we 
have calculated also the first moment
value of the structure function $g_1$ at different $Q^2$.
Using Eq.(\ref{5}), we recalculate the SMC measured asymmetry of the
proton and deuteron and E154 one of neutron
 at $Q^2= 100~ {\rm GeV}^2$, $30~ {\rm GeV}^2$,
$Q^2= 10~ {\rm GeV}^2$ and $3~ {\rm GeV}^2$
and get the value of $\int g_1(x) dx$ through the
measured $x$ ranges.  To obtain the first moment values
$\Gamma_1^{p(d)}$ we use an original estimations of SMC and E154
for unmeasured regions.
In Table we present the 
results for the mean values of $\Gamma_1^p - \Gamma_1^n $,
because the errors coincide with the errors of original analyses
\cite{SMC}-\cite{Q2E154}.
The value of $\Gamma_1^p - \Gamma_1^n $ at $Q^2$ = 10 GeV$^2$ obtained by 
direct DGLAP evolution are taken from article \cite{SMC}.\\

Let us now present the main results, 
following from the Table  
and the  Fig. 2.
\begin{itemize}
\item
The results are in very good
agreement with $g_1(x,Q^2)$ data of SMC, 
based on direct DGLAP evolution.
So, the suggestion about $Q^2$ independence of the  $A^*_1$ ratio
leads to correct results in simplest way. 
\item Our method allows to test of the Bjorken sum rule in a simple way 
with a good accuracy. 
Obtained results on the $\Gamma_1^p - \Gamma_1^n$ show that used experimental 
data well confirm the Bjorken sum rule  prediction.
\item
A violation of $A^*_1$-scaling (\ref{a2}) in (future) experimental data
will demonstrate
large gluon and/or sea quark contributions to SF $g_1(x,Q^2)$.
Then, the check of $Q^2$ dependence of $A^*_1$ ratio should be very useful
for an indication 
of large contributions
of gluons and sea quarks 
in future polarized experiments.
\end{itemize}

{\large{\it Acknowledgments.}}~~
One of us (AVK) is grateful very much to Organizing Committee of 
Workshop DUBNA-SPIN01
for the financial support. 
A.V.K. was supported in part by Alexander von Humboldt
fellowship and INTAS  grant N366.


\begin{thebibliography}{99}
\bibitem{AEL} M. Anselmino, A. Efremov and E. Leader,
Phys. Rep. {\bfseries 261}, 1 (1995).
%
\bibitem{Bj}
J.D. Bjorken, Phys. Rev. {\bfseries 148}, 1467 (1966); {\bfseries D1}, 1376
(1970).
%
\bibitem{EJ} J. Ellis and R.L. Jaffe, Phys. Rev.
{\bfseries D9}, 1444 (1974); {\bfseries D10}, 1669 (1974).
%
\bibitem{EG}
SM Collab., B. Adams {\it et al.}, Phys. Lett.
{\bfseries B329}, 399 (1994); {\bfseries B357}, 248 (1995); SLAC-E143
Collab., K. Abe {\it et al.}, Phys. Rev. Lett.
{\bfseries 74}, 346 (1995); {\bfseries 75}, 25 (1995); Phys. Rev.
{\bfseries D58}, 112003 (1998).
%
\bibitem{SMC}
SM Collab., B. Adams {\it et al.}, Phys. Rev.
{\bfseries D56}, 5330 (1997);  Phys. Lett.
{\bfseries B396}, 338 (1997).
%
\bibitem{E154n}
SLAC-E154
Collab., K. Abe {\it et al.}, Phys. Rev. Lett.
{\bfseries 97}, 26 (1997).
%
\bibitem{Q2E154}
SLAC-E154
Collab., K. Abe {\it et al.}, Phys. Lett.
{\bfseries B405}, 180 (1997).
%
\bibitem{EK93} J. Ellis and M. Karliner, Phys. Lett.
{\bfseries B313}, 131 (1993);  F.E. Close and R.G. Roberts,  Phys. Lett.
{\bfseries B316}, 165 (1993).
%
%
\bibitem{GRSV}
R.D. Ball {\it et al.}, 
Phys. Lett. {\bfseries B378}, 255 (1996);
M. Gluck {\it et al.}, 
Phys. Rev. {\bfseries D53}, 4775 (1996);
T. Gehrmann and W.J. Stirling, Phys. Rev. {\bfseries D53}, 6100 (1996);
A.V. Kotikov and D.V. Peshekhonov, hep-ph/9604269;
Phys. Atom. Nucl. {\bfseries 60}, 653 (1997);
Phys. Rev. {\bfseries D54}, 3162 (1996);
E. Leader {\it et al.}, 
Phys. Lett. {\bfseries B462}, 189 (1999);
S.~Simula {\it et al.},
Preprint RM3-TH/01-4 (hep-ph/0107036).
%
\bibitem{KP}
A. V. Kotikov and D. V. Peshekhonov,
JETP Lett. {\bfseries 65}, 7 (1997); 
{\it in} Proc. of the Workshop DIS96 (1996), p.612
(hep-ph/9608369). 
%
\bibitem{KNP}
A. V. Kotikov and D. V. Peshekhonov,
Eur. Phys. J. {\bfseries C9}, 55 (1999); 
{\it in} the Proc. of the Workshop DIS98 (1998), p.242
(hep-ph/9805374). 
%
\bibitem{Ermolaev}
J. Bartels {\it et al.}, 
Z. Phys. {\bf C72}, 627 (1996); {\bf C70}, 273 (1996);
B.I.~Ermolaev {\it et al.}, 
Z. Phys. {\bf C69}, 259 (1996);
Nucl. Phys. {\bf  B594}, 71 (2001); {\bf  B571}, 137 (2001);
{\it in} Proc. of the 
Workshop 
DIS2001 (2001), 
(hep-ph/0106317).
%
\bibitem{LoYn}
F. Martin, Phys. Rev. {\bf D19}, 1382 (1979);
C.~Lopez and F.I.~Yndurain,
Nucl. Phys. {\bf  B171}, 231 (1980);
A.V. Kotikov {\it et al.}, 
Theor. Math. Phys. {\bf 84}, 744 (1991).
%
\bibitem{Ko}
L.~L.~Jenkovszky {\it et al.}, 
Sov.\ J.\ Nucl.\ Phys.\  {\bf 55}, 1224 (1992);
A.V. Kotikov,
Phys. Atom. Nucl. {\bf 56}, 1276 (1993);
{\bf 57}, 133 (1994);
{\bf 59}, 2137 (1996);
Phys. Rev. {\bf D49}, 5746 (1994);
Mod. Phys. Lett. {\bf A11}, 103 (1996);
A.V. Kotikov {\it et al.}, 
Theor. Math. Phys. {\bf 111}, 442 (1997).
%
\bibitem{KPS1}
A.L.~Kataev {\it et al.}, 
Preprint CERN-TH/2001-58 (hep-ph/0106221);
V.G. Krivokhijine and A.V.~Kotikov, Dubna
preprint E2-2001-190 (hep-ph/0108224).
%
\bibitem{CCFRN} CCFR/NuTeV Collab., W. Seligman  {\it et al.},
Phys. Rev. Lett.  {\bfseries 79}, 1213 (1997);
W. Seligman, PhD Thesis (Columbia University), Nevis Report 292.
%
\bibitem{NMC} NM Collab., M. Arneodo {\it et al.},
Phys. Lett. {\bfseries B364}, 107 (1995).
%
\bibitem{SLAC} L. W. Whitlow {\it et al.},
Phys. Lett. {\bfseries B250}, 93 (1990).
%

\end{thebibliography}
\end{document}